# Flipping a Graduate-Level Software Engineering Foundations Course

Hakan Erdogmus  and  Cécile Péraire
Department of Electrical and Computer Engineering
Carnegie Mellon University, Silicon Valley Campus, USA
{hakan.erdogmus, cecile.peraire}@sv.cmu.edu

*Abstract*—Creating a graduate-level software engineering breadth course is challenging. The scope is wide. Students prefer hands-on work over theory. Industry increasingly values soft skills. Changing software technology requires the syllabus to be technology-agnostic, yet abstracting away technology compromises realism. Instructors must balance scope with depth of learning. At Carnegie Mellon University, we designed a flipped-classroom course that tackles these tradeoffs. The course has been offered since Fall 2014 in the Silicon Valley campus. In this paper, we describe the course's key features and summarize our experiences and lessons learned while designing, teaching, and maintaining it. We found that the pure flipped-classroom format was not optimal in ensuring sufficient transfer of knowledge, especially in remote settings. We initially underestimated teaching assistantship resources. We gradually complemented video lectures and hands-on live sessions with additional live components: easily replaceable recitations that focus on current technology and mini lectures that address application of theory and common wisdom. We also provided the students with more opportunities to share their successes and experiments with their peers. We achieved scalability by increasing the number of teaching assistants, paying attention to teaching assistant recruitment, and fostering a culture of mentoring among the teaching team.

## I. Introduction

Carnegie Mellon University's College of Engineering offers a graduate degree in Software Engineering in its Silicon Valley campus. Founded in 2014, the professional Master's program is hosted by the Department of Electrical and Computer Engineering (ECE) to cater to the needs of the local companies. The program admits students with a background in a software-related field or sufficient software development experience. It launched with 32 students in September 2014 and currently has over 160 full-time students. A core course in this program is Foundations of Software Engineering (FSE), whose purpose is to level the playing field for incoming students and serve as a gateway to more specialized software-focused courses. This paper describes the design of this complex breadth course using a flipped classroom format and documents our experiences teaching and maintaining it over multiple offerings. Currently in its sixth offering (with multiple sections), the course is still evolving. There have been multiple changes to its scope, content, and delivery. The course is taught twice a year (Fall and Spring) to 70-80 students per semester divided into two sections. The Spring offering has a multi-location remote section, concurrently broadcast to students in Pittsburgh, Pennsylvania, and Kigali, Rwanda.

Creating a graduate-level software engineering breadth course is challenging. The scope is wide and varied. Students come from various backgrounds, possess different skills, and have different levels of experience. They want to learn through hands-on work that will increase their job prospects and help them in the beginning of their careers. Changing software technology requires the syllabus to be technology-agnostic, yet abstracting away technology compromises realism and instant applicability. Instructors must manage several tradeoffs among variety of topics, technical skills, soft skills, depth of learning, rigor, practicality, flexibility, topicality, individual learning, team learning, mentoring, and scalability.

The paper is structured as follows. The next section reviews related work. We then describe the course's genesis and present its design, covering learning objectives, overall structure, team project, team formation, mentoring, and student assessment. Next we examine the structure of the course project's two-week long iterations. Finally, we reflect on the course's evolution over five semesters and discuss the lessons we have learned while designing, teaching, and maintaining it. The lessons address a variety of issues, from value of co-instruction, evolution toward a mixed-mode delivery, challenges of video creation, how to incentivize students to prepare for live sessions, how to scale to large audiences, distributed classrooms, role of teaching assistants, importance of peer evaluation, and more.

## II. Related Work

The *flipped*, or *inverted*, classroom is a technology-assisted pedagogical method that has been becoming increasingly popular [1]. It is based on flipping the delivery of theory and application. In a traditional classroom, theory is provided during in-class lectures. Students apply the theory outside the classroom through take-home assignments. In a flipped classroom, this process is reversed: theory is provided before class through online pre-recorded lectures and supporting materials. During the in-class session, which we refer as a *live session*, students apply the theory through supervised activities.

While the merits of flipping a classroom at lower levels of education are still controversial among educators, the method is generally thought to be beneficial in higher education settings, where its benefits allegedly increase with the maturity of students [2, 3]. Advantages include optimization of class time, improved engagement by students, support for abstract and conceptual thinking, and enhancement of instructor-student and student-to-student interactions. Disadvantages include increased instructor effort for preparation of online materials and class activities and difficulty in sustaining student motivation to perform both preparatory work before attending live sessions and reinforcing work after attending them [4].

In computer science and software engineering, the reported uses of flipped classroom have mostly been limited to undergraduate education and introductory programming courses. Hayashi et al. [5] observed increased performance outcomes for programming courses in a three-year study comparing the flipped classroom approach with a traditional method. Moore and Dunlop [6] also report improved outcomes, this time in a graduate computer science course. However individual experiences still vary in computer science undergraduate education [7], and failures have been reported [8]. A study by Horton et al. [9] reported that while exam performance increased with flipped classroom, student enjoyment was unaffected. Another study [10] associates performance with grade-based incentives, with variable results when grade incentives are removed. Köppe et al. [11] document some common non-grade-related patterns that have proven effective in flipped classrooms.

Reported uses in software engineering curricula are scarce: Kiat and Kwong [1] recount positive experiences on student motivation in a pilot module of an undergraduate course. We have not found any accounts of flipped classroom use in a broad-scope software engineering graduate course, where concepts tend to be more complex multi-faceted, and abstract compared to undergraduate education, and classroom activities typically require higher-order thinking and more intense collaboration.

### III. COURSE DESIGN

#### A. Background and Process

The general scope of FSE was determined by a departmental ad-hoc committee prior to the launch of the new graduate software engineering program. The committee report included a long list of over 30 topics that covered most areas of the Software Engineering Body of Knowledge (SWEBoK) [12]. To reduce and fine-tune the scope and decide on the final syllabus, we first reviewed SWEBoK and existing US-based software engineering graduate programs. We also analyzed the content of previous software engineering courses offered at our campus. Most importantly, we conducted 18 interviews of Silicon Valley employers as well as half a dozen ECE faculty and several existing CMU Silicon Valley students to understand the needs of our key stakeholders. These efforts helped us identify and prioritize teaching objectives and course topics. The resulting syllabus consisted of 11 main modules organized around seven iterations and corresponding themes of a semester-long team project.

With encouragement from the department and motivated by optimizing class-room time and increasing student interaction, we settled on the flipped classroom format as the main method of delivery. In the summer of 2014, we attended a five-week-long online workshop provided by Acatar, a CMU spinoff specialized in technology enhanced learning. We also designed the theme project and hired a student team to pilot it. During the workshop, we learned how to create an effective flipped course using the Acatar's in-house learning platform. With help and advice from Acatar experts, we then proceeded to implement the syllabus by developing the instructional videos, supporting materials, and live session activities. Videos were recorded using the Panopto (panopto.com), ScreenFlow (telestream.net) and professional-grade audio equipment. We estimate the total initial effort involved in preparing the course during the summer of 2014 at about 1000 person-hours by two faculty members and a team of three students.

#### B. Learning Objectives

In the syllabus, we defined the learning objectives as follows:
- compare and reflect on different approaches and methods for engineering software systems and combine them to fit a specific context;
- apply and interleave fundamental requirements, architecture, design, construction, and testing techniques;
- leverage contemporary development tools and environments to boost productivity and maintain quality;
- demonstrate working familiarity with central modeling concepts and notations;
- plan, estimate, and manage a software project in an incremental and iterative fashion; and
- using contemporary software engineering practices, develop and deliver a realistic software project that meets stakeholders' expectations, is high-quality, and balances underlying engineering tradeoffs.

Notable about the list is its inclusion of both soft and hard skills and equal emphasis on both engineering and management concepts.

#### C. Overall Structure

The course is spread over 14 weeks, with twice weekly 110-minute live sessions. It is designed to require on average 12 weekly hours of student effort, including preparation for live sessions, contact hours, team project components, and assessments. Course outline of a past offering is given in Table I. Course modules are aligned with project iterations, each of which focus on a distinct software engineering discipline, starting with *Teamwork & Technology*, followed by *Architecture & Design*, *Construction, Testing & Quality*, and *Requirements*. The final iteration focuses on *Closure & Presentation*. Guest lectures are added to provide students with variety and outside perspectives.

#### D. Team Project

Since FSE aims to bring all incoming students to the same level before they take specialized SE courses, we decided to limit the available degrees of freedom by opting for a highly streamlined, semester-long team project with both fixed and student-defined requirements. Earlier iterations of the team project are fixed to focus on practices. Later iterations can be shaped by the students individually and/or project teams to encourage creativity and maintain motivation.

We opted against an open-ended project to standardize learning, which was a priority in a breadth course with numerous moving parts. We also decided to fix the programming languages, but leave other technology choices to the teams, subject to certain constraints.

TABLE I. OUTLINE OF THIRD FSE OFFERING.

| Wk | Theme | Project | Live Session 1 | Live Session 2 |
|---|---|---|---|---|
| #1 | Introducing Software Engineering | N/A | Course Introduction | Activity: Combining Scrum & Kanban |
| #2 | Managing Your Software Project | Iteration 0 - Teamwork & Technology | Project Planning, Estimation, Measurement, and Tracking | Intro to Project and Iteration 0; Activity: Planning Poker |
| #3 | Working with Objects and Models | | Activity: Using UML to Model Scrum | Project Reflection |
| #4 | Mapping System Behavior to Design | Iteration 1 - Architecture & Design | Activity: Object-Oriented Analysis | Demo of Iteration 0; Intro to Iteration 1 |
| #5 | Seeing the Big Picture & Detailing the Design | | Activity: Design Patterns | Activity: Architecture |
| #6 | Ensuring Steady Progress with Technical Practices | Iteration 2 - Construction | Guest Lecture: Software Engineering Practices | Demo of Iteration 1; Intro to Iteration 2 & Unit Testing Lab |
| #7 | Reasoning about Technical Debt | | Activity: Technical Debt | Mid-term Quiz |
| #8 | Making Sure Software Works, Mostly | Iteration 3 - Testing & Quality | Activity: UI Testing | Demo of Iteration 2; Intro to Iteration 3 |
| #9 | Achieving High Quality | | Activity: Code Review | Project Reflection |
| #10 | Capturing What Software Should Do | Iteration 4 - Requirements | Activity: Requirements Engineering | Demo of Iteration 3; Intro to Iteration 4 |
| #11 | Managing Computing Resources | | Guest Lecture: Managing Computing Resources | Project Reflection |
| #12 | Reasoning About Quality Attributes | Iteration 5 - Putting It All Together | Activity: Power Lab | Demo of Iteration 4; Intro to Iteration 5 |
| #13 | Course Review | | Course Review Q&A | Demo of Iteration 5 |
| #14 | Presenting Your Project Solution | Iteration 6 - Closure & Presentation | Project Video Presentations | |

A preparatory coding assignment is sent to all incoming students at least one month before the start of classes to ensure baseline familiarity with central technologies needed in the project, such as programming languages (e.g., *node.js* and the web stack), frameworks (e.g., *express.js*), and tools (e.g., *git* and *github*). Students are not admitted to the course

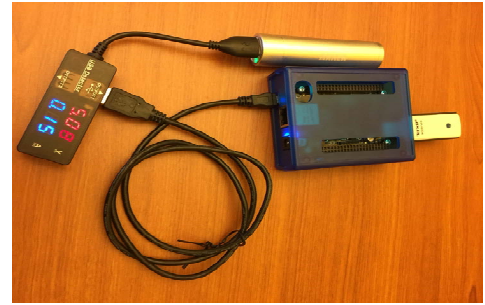

Figure 1. Team project hardware kit used in first four

if they cannot successfully complete the preparatory assignment.

The project is executed in multiple iterations that build on each other. In fact, the course and the flipping structure revolve around the project iterations. Each iteration has a distinct theme and focuses on a software engineering discipline and associated practices. Weekly live session activities are designed to support these practices. Starting in Iteration 1, each iteration results in working software. The iterations are sequenced in the style of the Rational Unified Process [13] to emphasize different disciplines and associated practices as the project matures, as shown in Table II. The table shows functional and practice requirements associated with each iteration. Notably, the *Requirements* iteration is the theme of the next to last iteration to perturb the traditional waterfall sequence and convey that requirements-related activities do not always come first. To reinforce this idea further, we provide the fixed-requirements of each iteration in a piecemeal manner: certain requirements in later iterations deliberately affect decisions made in previous iterations and trigger major refactoring. This design feature invariably stir lively discussion among students.

Originally, the team project (dubbed *Survivable Social Network on a Chip*, or SSNoC, in the first four offerings) was inspired by a previous CMU research initiative on emergency management. We chose a hardware kit as the deployment platform, consisting of a system-on-chip board to host a self-contained server application, a wireless dongle capable of serving as a Wireless Access Point for clients, a rechargeable battery for self-sufficiency, and a USB power meter to take energy consumption measurements (see Fig. 1). This platform choice allowed us to bring a systems perspective to the project and the course, impose resource constraints, and address resource management in an applied context. It also helped align the course with hardware and systems-oriented ECE courses. In the last offerings, however, we removed the hardware kit in favor of cloud deployment (more on this later).

*E. Team Formation and Mentoring*

Project teams of four to five students are formed early in the course using a team formation tool called CATME [14]. Students are asked to fill out an online demographic survey on their background, relevant knowledge, skill level with specific technologies, orientation, leadership preferences, gender, and ethnicity. We then tune the tool's parameters to achieve the desired level of diversity and balance. Teams are formed automatically by the tool. Manual adjustments are made before the teams are finalized.

Each project team is assigned a teaching assistant (TA) who mentors the team and monitors its performance and dynamic. TAs lead in-class reflection meetings and play the role of the product owner during demo days. They also meet their teams outside live sessions on an as needed basis.

*F. Student Assessment and Evaluation*

55% of a student's final grade is based on individual performance, which is assessed through (1) two formal online quizzes (midterm and final); (2) a graded lab conducted as a live-session activity; (3) an individually graded project iteration performance (added in last two offerings); and (4) participation. Participation is assessed through attendance to live sessions (which is mandatory), activity in live-session discussions, activity in assessment polls and Q&As, interaction with the teaching team, meaningful contributions to the online discussion board, activity during reflection and demo periods, and volunteerism in show-and-tell presentations (added in last two offerings). The goal of the participation grade is to encourage students to become active learners.

45% of the student's final grade is based on project team performance. However, this grade is individually adjusted via 360-degree, anonymous peer evaluations conducted using the CATME tool. Each team receives a team score based on their cumulative project performance and end-of-project team presentation. This score is adjusted upward or downward by a multiplier derived from the peer evaluations.

The majority of the students perform reasonably well during the course, with a failure rate (C or below) of about 10%. Between 10-15% of students perform at the highest level.

## IV. ANATOMY OF A FLIPPED ITERATION

Each iteration of the course lasts two weeks and structured around the following components: (1) brief presentation of the iteration objectives; (2) *concept-based live sessions* where students learn by applying the underlying concepts necessary to reach an iteration's goal; and (3) *project-based live sessions* where students reflect on their team work and demonstrate their results.

*A. Beginning of an Iteration*

An iteration starts with a brief introduction by the instructor explaining the iteration's theme, functional requirements to be implemented, practices to be introduced, and deliverables expected. A checklist akin to a *Definition of Done* list is provided, as illustrated in Table III. One to three new use cases are assigned at each iteration; however, the use cases are not discussed in detail in class (the teams review the use case specifications offline).

*B. Concept-Based Live Sessions*

These live sessions (Live Session 1 column in Table I) are typically dedicated to an in-class team activity related to the iteration theme and the practices to be applied. Students prepare for the live session by watching the assigned video lectures and reviewing other supporting resources. Video lectures cover modules that last between ten to 25 minutes. All video lectures are professionally captioned to allow the students another option to follow the speaker's narrative. Convenient navigation features help locate desired sections, as seen in Fig. 2. Fig. 3 shows example screenshots for a typical week's agenda.

TABLE II. SAMPLE TEAM PROJECT STRUCTURE AND ITERATION REQUIREMENTS

| Iteration | New Functional Requirements | New Practices |
|---|---|---|
| 0. Teamwork & Technology | • None | • Team Charter Definition<br>• Technology Selection<br>• Collaboration Tools<br>• Estimation & Planning<br>• Reflection |
| 1. Architecture & Design | • Join Community<br>• Chat Publicly | • Architecture Definition<br>• Object-Oriented Analysis and Design |
| 2. Construction | • Share Status<br>• Chat Privately<br>• Post Announcement | • Pair Programming<br>• Version Control<br>• Continuous Integration<br>• Unit Testing |
| 3. Testing & Quality | • Search Information<br>• Measure Performance<br>• Measure Memory | • UI/Acceptance Testing<br>• Code Coverage<br>• Code Review<br>• Static Analysis |
| 4. Requirements *(individually graded iteration)* | • *Requirements defined individually by each student* | • Use Case Definition<br>• UI Mockup Creation |
| 5. Putting It All Together | • Administer User Profile<br>• *Requirements defined by the team* | • User Story Definition<br>• Refactoring |

Upon physically joining a concept-based live session, the students log in to the course's *AdobeConnect Pro* (ACP) site for in-class assessment. The assessment is performed through ACP polls that cover the assigned materials. Two to five polls are released by the instructor as multiple-choice or short answer questions, often with cues to guide students. All students are required to respond to the polls individually on ACP. The answers are then revealed and sample solutions are discussed as a group.

Q&A assessment via polls allows faculty to evaluate the student's level of understanding of the material and clarify confusing points, further preparing the students for the live session's activity. The live session then proceeds to the in-class activity. The instructor introduces the rules and steps. Sometimes this portion includes a mini-lecture addressing the subtle points of theory. Project teams stay together during the activity and are monitored and guided by the teaching team. The teaching team circulates through the room to answer questions, encourage collaboration, and help the teams when they get stuck. At the end of the activity, selected teams present and discuss their results. The live session ends with the teams electronically submitting their activity artifacts.

TABLE III. PARTIAL SAMPLE CHECKLIST FOR A NEW ITERATION

| Element | Criteria |
|---|---|
| Project / Iteration Plan | ❑ Updated Trello board. |
| Continuous Integration (CI) Demonstration | ❑ Make a trivial change to the code. Push code and show that the CI server successfully builds it (and run any unit tests). |
| Use Case: Share Status - Working System | ❑ Work is DONE. System behavior FULLY satisfies use case specification. It is deployed in the cloud and demoed on the phone.<br>❑ RESTful API for use case is functional & documented.<br>❑ Unit tests for server-side code are created, pass, and cover the code below the level of the RESTful API. |
| Practices | Practices are in use (**bold**: mandatory):<br>❑ **Estimation**<br>❑ Reflection<br>❑ **Architecture Definition**<br>❑ OOAD (as needed)<br>❑ Design Patterns (as needed)<br>❑ **Pair Programming**<br>❑ **Continuous Integration**<br>❑ **Continuous Deployment**<br>❑ **Unit Testing** |

*C. Project-Based Live Sessions*

The second live session of the week is project-based (Live Session 2 column in Table I) and typically alternates between a mid-iteration *reflection* (concluding the first week of an iteration) and an iteration *demo* (concluding an iteration).

During the *reflection period*, project teams work with their TA and reflect on their progress by conducting two types of activities. The first activity is a standard retrospective-style [15] open discussion focusing on what has been working well, what has been problematic, and what needs improvement since the last reflection period. The teams follow up on any action items outstanding from previous meetings and identify one point for improvement to focus on until the next reflection period. The team records the point as an action item on it project board.

The second activity is a classical *Scrum*-style short standup meeting [16] during which each team member answers three personal questions: what the team member has done since the last reflection period, what the team member is planning to do until the next reflection period, and whether the team member has any blockers that require assistance. The teams are directed not to try to solve technical problems during these meetings, but record them on their project boards for later resolution.

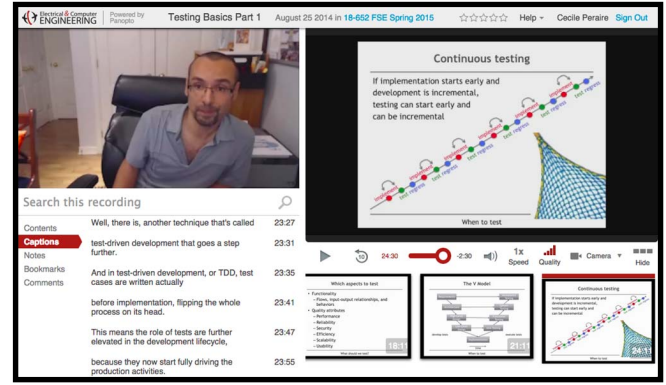

Figure 2. Video lecture with captions pane (bottom left), lecturer (top left), visuals (top right), and navigation (bottom right).

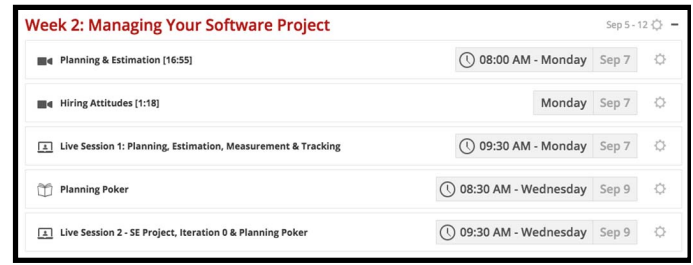

Figure 3. Agenda for a typical FSE week.

The *demo period* takes the form of an acceptance meeting where the TA systematically goes through each requirement and deliverable of an iteration with the team using the checklist for that iteration. The teams are asked to demonstrate the functionality of their applications using the deployment configuration. Each functional requirement is carefully reviewed for compliance with the associated use case specification. Required practices are discussed. Each artifact on the deliverables list is checked. The TAs record any outstanding items on a team spreadsheet with notes and feedback. The spreadsheet is accessible to the team. Any outstanding items are revisited during the next demo period.

*D. Deviations from Standard Structure*

Although iterations typically follow the standard structure described above, variations have been implemented over time. These variations include mini live lectures, individual activity in a live session, tech talks, and show-and-tell presentations. These additions are discussed under lessons learned below.

V. COURSE EVOLUTION AND LESSONS LEARNED

The course has constantly been evolving since its first offering. We have performed many experiments. Not all of them have been as successful as we had hoped, forcing us to make small and big changes along the way and update the next offering. Occasionally, a feature that works in one offering is not as effective in a subsequent offering because of a subtle shift in student demographics, technologies used, or environment. Below we summarize the lessons learned within their respective contexts, and when applicable, accompanied by changes affected in response to those lessons learned. When applicable, we also relate the lessons learned to flipped classroom patterns documented by Köppe et al. [11].

### A. Co-Instruction

Co-instruction is one aspect that has consistently worked well in FSE. The amount of effort involved in creating, delivering, and implementing a graduate breath course as a flipped classroom with a well-structured and tightly-managed project component is quite high. We found the division of work to be essential in such an endeavor. The scope is wide with many topics to cover, but insights and learnings still need to be deep enough. The course has greatly benefited from our complementary expertise. Additionally, we always need an objective pair of eyes to provide feedback to one another and check each other's work. We found this to be invaluable in making improvements. Co-instruction with tight collaboration has therefore been pivotal in the success of the course.

### B. Pure Flipping vs. Mixed-Mode Delivery

The first offering of FSE was close to pure flipped-classroom with no in-class lectures. Video lectures free up time during live sessions. Students are able to view a video at their own pace, rewinding and re-watching sections as they please. Our video lectures are fully captioned, which allows some students to follow them more easily. However, there is no rapport or eye contact with the students during a video lecture. Instructors cannot receive immediate feedback, or gauge the level of attention and comprehension from the viewer's body language or social cues. Reciprocally, students cannot ask impromptu questions to the instructor for immediate clarification. By the end of the first offering, we started to feel uncomfortable with the lack of interaction, intervention, and control. We also realized that students consistently misunderstood certain abstract concepts or got certain subtle points in the application of the theory wrong. These misconceptions were hard to correct with just Q&A style live assessment. Knowledge transfer was not adequate and entirely effective with only video lectures.

*Complement recorded lectures with mini live lectures.* Starting with the second offering, we gradually introduced mini-lectures lasting between ten to 20 minutes, rolled into live session Q&As and activity introductions. This addition allowed us to go deeper into certain subjects and amend the content as we deemed appropriate from offering to offering. It provided instructors and students with the opportunity to discuss, clarify, and emphasize concepts insufficiently addressed or missed in recorded lectures. We strive to keep the mini-lectures as interactive as possible. We think that the ability to amend the content is particularly important in a software course, where the pace of change is high.

*Add tech talks to cover technology.* We also added TA- and instructor-led recitations, or *tech talks*, that are essentially technology-focused short tutorials, demos, or experience-based presentations. The tech talks focus on frameworks, languages, and tools used in the team project, and are subject to change each semester. Students value them greatly. Driven by demand, these talks take place during the first few weeks of the course. Depending on their length, they are scheduled either outside regular class hours or at the beginning of a demo or reflection period.

Both mini lectures and tech talks follow the pattern "Add Value Beyond Feedback" described by Köppe et al. [11]. In this reference, the authors state that in a flipped classroom course, "class-session is spent primarily on giving students feedback on their individual homework, students might perceive the session as not very valuable, reducing both the effect of the session and their motivation to prepare diligently for coming sessions." Their advice is to "interweave feedback with added value moments: mini-lectures with new knowledge, interesting demo's, anecdotes with examples from real-life, generalized wisdom etc."

Our current offering can thus be said to be a mixed-mode course rather than purely flipped. We find this mode to be more effective for our purposes as it leverages the strengths of both traditional and contemporary pedagogical techniques.

### C. Video Creation

Producing videos is extremely time consuming, but an opportunity to improve teaching and presentation skills when done with the guidance of a mentor. We do not think we have mastered this skill, but we learned a great deal from experts who mentored us. Below is advice that helped us with this process:

*Keep them short and focused.* Videos should be created to retain students' attention and maximize learning: they should be kept short and convey a limited number of concepts and key messages. The key messages should be easy to summarize at the end. We had trouble keeping the videos concise at first. Students were not shy about complaining. Later we broke up long videos into shorter bits.

*Include required elements.* Elements that should be included in a video include a (catchy) opening with motivation, agenda, learning objectives, and summary of key messages.

*Pictures over text.* Prefer graphics and pictorials over text in visuals.

*Ask for participation.* A video lecture may encourage active participation of the viewer, for example it may pose a question and ask the viewer to pause and ponder the question or solve a problem.

*Principles over fashion.* Videos should focus on principles and foundational concepts versus technology and fads to maximize their relevance in fast-evolving subjects. Keep timeless components in; remove volatile components that are likely to become stale.

*Stabilize before recording.* Video lectures should ideally be created once the content has been tested and stabilized. Unfortunately, we could not follow this advice. We were designing the course almost entirely from scratch, and took many risks with untested content. We later had to revisit and edit existing videos to make changes (which was extremely time-consuming). We also had to eliminate content that did not work. Be prepared to rework or trash some portion when designing a flipped classroom from scratch.

### D. Student Preparation

*Tool support is important for controlling pace.* Our primary tool, the Acatar Learning Environment (ALE), was instrumental in managing the pace of the course and communicating with the students, although we had to replace it in 2016. Students demanded clear instruction on how to prepare for live sessions and when each preparation element was due. ALE provided the necessary scaffolding for weekly requirements with clearly identified deadlines. The students were automatically taken to the current or impending week's agenda upon logging in and pace themselves accordingly. Köppe et al. [11] capture this behavior in a flipped classroom pattern called "Controlled Pace," advising instructors to "control the pace of the students through an explicit planning per in-class meeting and deadlines for submitting the preparation prior to the in-class meeting."

*Incentivize students.* In live sessions, we implement assessment mechanisms to encourage students to perform the required preparations (watch assigned videos, set up tools, read mandatory readings). These mechanisms include informal quizzes, online polls (released via the ACP tool), and free-form Q&A sessions. Assessment activities take place at the beginning of each live session. They are not individually and explicitly graded, but students receive points that count toward their participation grade. Preparation assessment is central to the flipped classroom. When we skipped these mechanisms, we consistently observed a notable drop in student preparedness. As a result, live session activities were not as successful in those cases. Unprepared students have a substantial negative impact on live session outcomes. Students require appropriate incentives to prepare. Even though our incentivization strategies work well with most students, they fail with a subpopulation who still come to class unprepared. We have not completely figured out the subtleties of incentivization.

### E. Importance of Learning Objectives

Learning objectives are a powerful mechanism for driving syllabus design at all levels (course, module, activity, and lecture). While creating the course, our mentor placed a great deal of emphasis on learning objectives [17]. This emphasis was particularly useful in both video lecture and live session design. One advice was about making learning objectives *active*, by choosing active and concrete verbs over passive and abstract verbs, for example "design," "apply," "use," and "list" over "understand," "learn," and "characterize." This advice helped us to think of a course component in terms more actionable by the students, and likely increased the effectiveness of the component.

### F. Effective Live Sessions

Live session activities allow students to become active participants of the learning process rather than remain passive listeners. Students can *learn by doing*, reflect on their work, and share their solutions.

*Leverage student solutions for learning.* Köppe et al.'s "Compare Solutions" pattern [11] posits that that "students are only familiar with their own solution, and are unable to recognize strengths and weaknesses in them." They suggest "showing multiple solutions to the same problem that are comparable and differ in interesting ways." We consistently apply this pattern in live session activities by having selected student teams present their solutions and compare them to other teams' results. We encourage students to comment on how their solutions differ and why. We also learn from these interactions: occasionally students come up with interesting, novel, worthwhile ideas that we had not considered. We incorporate the best ideas to live session discussions in future offerings.

*Encourage students to share.* Since the second offering of the course, once the students are well into their team projects, we ask teams to volunteer to share their trials, tribulations, and successes in short *show-and-tell* presentations (limited to 5-10 mins) followed up with a Q&A and discussion session. Teams have the option to talk about a technology or practice they have mastered, demo a tool that they think might be useful to others, or share an experiment and lessons learned. We offer participation points to the presenters in return. We do not think that students take the offer in exchange for points: they like to share their unique achievements and learnings for peer recognition. Almost every project team volunteers once they see other teams present. Peer pressure works well here. This strategy mirrors the "Use Student Solution" and "Every Student Solution Counts" patterns described by Köppe et al. [11].

*Provide ample opportunities for reflection.* Reflection is a cornerstone of experiential learning [18]. In addition to bi-weekly reflection periods in live sessions, project teams are encouraged to hold reflection meetings on their own outside of live sessions. Reflection periods help the teaching team identify dysfunctional behavior, and we re-use the retrospective format in any subsequent offline intervention. Reflection also happens implicitly during Q&A sessions, mini lectures, student presentations, and polls by prompting students to explain why they think a particular idea or solution works.

*Avoid overuse of tools.* While software engineering is a tool-intensive field and hands-on learning often requires tools, we found that excessive tool usage hinders learning by focusing the students' attention on the tool rather than on application of concepts and skills. For example, during a hands-on version control activity using *git* and *github*, students focused excessively on *github* features and *git* syntax. This unanticipated shift in attention came at the expense of learning about useful version control workflows. In the last offering, we had plans to integrate *github* pull requests [19] into the code review team activity, but decided against it for fear of repeating the version control experience. Instead the TAs provide a demo of pull requests during a recitation, which the students have the option of following through using their own laptops. This recitation is recorded. After the code review activity, the students are asked to review the recitation video to remember how to leverage *github* pull requests in code reviews.

### G. Scaling Challenges and Mentoring

*Plan succession of TAs, mentor them, and empower them.* We designed live session activities and the team project to maximize both student-to-student and student-to-instructor interactivity. However, we needed to also scale the course to up to 80 students per offering. This posed a challenge. With so many interacting parts, we could not do this as two instructors no matter how much streamlined the course became, and we desperately needed capable TAs. We were lucky with TA resources in that we could always find skilled and technically competent TAs. However, as enrolment grew, succession planning and training of TAs became more and more important, shifting our focus on role modeling, mentoring, and recruitment. We find that high-performing students naturally mimic our behavior, both teaching and mentoring style. Reflection meetings help us set an example for these students. Currently, we more proactively identify and recruit high-performing students with natural leadership and good social skills. We start mentoring recruited TAs at the beginning of the course via regular meetings in which the TAs and instructors together review each project team, and discuss characteristics of high-performing teams, issues with low-performing students, common student misunderstandings, and students' difficulties with and perceptions of the content and delivery. They practice their mentoring skills during reflection and demo periods. We invite the TAs to observe our interventions with dysfunctional teams when appropriate. By the middle of the course, they are often ready to lead. At that point, we empower them to intervene in problematic situations while monitoring them. We also encourage the TAs to collaboratively propose and develop instructional materials in the form of tech talks. The TAs invariably take pride in their work and consider their job as an important contribution to their own personal growth.

Our TAs receive a salary and are expected to spend an average of ten hours per week. Each TA oversees the work of two to three teams. The ideal ratio is one TA for two teams to give the TA enough time for effective mentoring.

*Do not mixed remote and local students.* Another complication of the course is the remote offerings, which impede face-to-face interaction in live sessions. Initially we mixed remote and local students in the same session. Our videoconferencing setup was adequate, but we could not observe the remote and local students with equal attention. Our attention naturally shifted to either local students who naturally tended to be more participative or to remote students who tended to be quieter. One group of students often felt neglected, and this sentiment was reflected in their course evaluations. We also felt like we couldn't manage the remote room as effectively as we wanted. We now have separate remote and local sessions, which allow us to focus our attention on one group of students at a time. We also make sure there is a remote TA or proxy instructor who can help manage the room in the remote location.

### H. Team Project

It is important to give the students a realistic practical component and develop their collaboration and teamwork skills. We designed the practical component as a semester-long iterative team project that employs a mix of traditional and modern software engineering practices.

*Maximize team interaction opportunities.* The flipped classroom provides opportunities for teams to work together during live sessions. We capitalize on these opportunities in regular reflection meetings, demo periods, and numerous team-based in-class activities during which students together learn and practice a technique they are asked to employ in their projects.

*Use theme-based iterations.* Assigning a theme to each iteration allows for a modern iterative development experience while focusing on the foundation at the same time. This feature of the team project has consistently worked well by aligning theory with practice.

*Overemphasize process.* Students naturally focus on concrete deliverables, which involve churning code that satisfies functional requirements. With bi-weekly milestones and demos that require working code, we had major challenges having students focus on software engineering process--the practices--versus implementation only. We gradually started to add graded assessment elements that explicitly relate to process concepts. These elements are now integrated into the iteration checklists and closely monitored by the TAs. We also started to overemphasize process in Q&A sessions, discussions, and mini lectures. Process has been a pain point from the beginning. While we are not entirely happy yet and dislike too much focus on grades, we have achieved reasonable success through explicit grade-based incentives.

*Standardize learning while allowing for creativity and experimentation.* The first offering provided the project teams with a skeletal application in the hopes of speeding up ramp-up and serving as a starter model. This strategy was a failure, and we abandoned it in the second offering. Students instead fixated on one solution, did not quite understand why it was chosen, and refused to consider alternative solutions. We still believe that the team project should not be open-ended: its purpose is not to double as a practicum with a real external client, but offer a realistic, yet controlled, experience, very much like an MBA strategy game. We experimented with different degrees of freedom through multiple offerings. We have finally settled on the following strategy that we believe works best (current selections are in brackets): fix the programming language (the web stack and server-side *JavaScript*), high-level architecture (client-server, REST, and sockets in a specific combination), main language framework (*express.js*), 75% of functional requirements, deployment configuration (server deployed in the cloud with a mobile browser client), and central tools (*github*, *Trello*); allow flexibility in vision (non-functional requirements and 25% of functional requirements), code organization (subject to design principles), helper frameworks, persistence layer, user interface, personal development environment, and other tools and services.

*I. Scope Challenges*

*Less is more.* There is one lesson we keep learning over and over again: less is more. However, the temptation to add new components or implement a "cool" idea is forever present. We now have the rule that when we add a component, we must give up another. In the first offering, the scope was too wide. The team project used multiple programming languages (both *Java* and server-side *JavaScript*). Teams had to deploy their application on a hardware kit. They had a second project on designing a software engineering process for a specific application domain. They had to prepare polished video presentations. There were extra modules, activities, and assignments that we eliminated over time as we introduced more specialized courses into the Master's program. In the second offering, we removed the second project and switched to a single programming language. In subsequent offerings, we eliminated some modules and videos that were not providing as much value as we had hoped. We used any slack to go deeper on remaining subjects. Despite the many optimizations implemented, the scope remains intentionally wide as FSE is centrally a breadth course. Depth is achieved outside of FSE through specialized courses that build on the foundations.

*Let go of your baby.* In the last offering we removed the hardware kit and associated components (a lab, hardware- and performance-related use cases, and a guest lecture) to reduce overhead and make space for an additional practice. The removal of the hardware kit was a tough decision: we had invested so much in making it work, but the overhead of maintaining, replacing, and preparing the kits became a real problem. Student feedback also showed it was not providing as much value as we had hoped beyond being a novelty. TAs and students wanted to know about *Continuous Deployment* as a natural next step to the existing practice of *Continuous Integration*. Novelty features are appealing for differentiating a course, but they may not indefinitely provide sufficient added value if their overhead is large and benefits are marginal.

*J. Learning Technology and Tools*

*Be ready for changes.* Learning technology software is rapidly evolving. After five semesters using our original learning management system, we had to urgently migrate our content to a new platform, Canvas (www.canvaslms.com), because the parent company of the original system had been acquired and that system was no longer available. It is wise to package the content so it is easy to recover the content and migrate to a new system.

*One-size does not fit all.* Our current offering uses a variety of technological aids on top of our main learning platform Canvas. These include *Piazza* for online discussions; *Google Drive* for shared dynamic content; and ACP for supporting live sessions. Over time, we had to admit that while our original learning management system was very good at providing the necessary scaffolding, it did not handle student submissions, assessments, and grading well. Unfortunately, each tool is often good at one task. While the proliferation of tools is not desirable and causes a certain amount of confusion, we find that students are quick to adapt. Many good tools are far preferable to poorly used tools with suboptimal functionality and missed interaction opportunities. As a case in point, our original system provided a discussion board feature, but students refused to use it because of its lack of notification mechanisms. When we switched to *Piazza*, the situation completely reversed and the discussion board became a central focus of the students outside live sessions.

*K. Grading Strategies*

*Balance team and individual performance.* The team project is effort intensive and has a large contribution to the final grade. This amplifies the impact of *free loading*, or insufficient contribution to team output. Even though teams are closely mentored and monitored, free loading continues to be an issue. To alleviate it, in the third offering, we converted one project iteration to an individual iteration, during which each student independently proposes one or more use cases and fully implements them. Students are graded individually for this part and they receive plenty of advance warning that they will be. The strategy has worked well since it prevents students from over-specialization, sticking to non-technical roles, and avoiding certain tasks. Additionally, one live session is performed individually rather than as a team and involves the student performing a lab based on a technical practice (currently user interface testing) using a tool on his or her laptop. This activity has a formal deliverable and is explicitly graded unlike most other team-based live session activities.

*Perform peer evaluations.* From the beginning, we have been conducting anonymous formal 360-degree peer evaluations at the end of the course using CATME, a research-based instrument [14]. Again, students are given heads up to disincentivize them against free loading. This strategy is successful in differentiating between high- and low-performing students in the same team. The instrument flags gaming patterns and teams with internal conflict. Peer evaluation results are triangulated with teaching team observations before project grades are finalized. Students in the same team may receive drastically different grades due to the impact of peer evaluation scores.

*Keep grading granularity coarse while maintaining visibility, feedback, and objectivity.* From the beginning, we have been averse to motivating students solely based on grades. While graded assessment is inevitable in our context, we have over time switched from a more granular to a coarser grading scheme, replacing intermediate grades with ungraded and feedback-based assessment. For example, in the first two offerings, each project iteration was graded separately at the end of the iteration. This focused the students overly on grades, prompting them to negotiate with the TAs during demo periods. Complaints were also voiced in course evaluations. Now we only maintain a checklist with constructive feedback, which is visible to all students in all teams. Teams can compare their progress to that of other teams. The project/iteration checklists are frequently updated. Teams also get a mid-semester progress report on their project. We conduct extra reflection meetings with low-performing teams following the progress report. These changes proved relatively successful in reversing the students' tendency to latch onto grades.

## I. CONCLUSIONS AND OPEN ISSUES

This paper described the design of a complex graduate software engineering breadth course that uses a flipped classroom format. It documented our experiences teaching and maintaining the course over multiple offerings. We offered numerous insights, including the value of co-instruction, evolution toward a mixed delivery format (flipped and traditional), tips for video lecture creation, use of tool support for team formation and peer evaluation, and need to keep grading granularity coarse. The current version of the course is substantially different from the initial offering in many aspects. The evolution, however, is not over. We are still on our journey of continuous improvement.

In a flipped classroom, students must be strongly and repeatedly encouraged to prepare for live sessions by viewing the assigned videos and reading the assigned materials, as well as ask clarification and probing questions. We are still looking for effective non-grade-based strategies for incentivizing students to come to class better prepared.

In the context of project-based learning, mechanisms encouraging students to focus on engineering processes are essential. Otherwise, students tend to focus predominantly on building the final solution while ignoring many facets of the development process, facets they need to master to effectively compete in the job market. We are still experimenting with ways of encouraging students to pay attention to the process, not only to the code.

Grading and free loading are challenges associated with team-based projects. We are still trying to find fairer ways of assessing individual performance while encouraging better collaboration among students.

Scaling-up our model in terms of number of students and geographic locations has been difficult. It requires relying on technically competent teaching assistants with excellent leadership skills. We are always interested in new strategies for improving the training of our teaching assistants and fostering a culture of mentoring.

## ACKNOWLEDGMENTS

We would like to thank Jelena, Kovacevic and Diana Marculescu for supporting our efforts and encouraging us to experiment with technology-enhanced learning. Marie Norman provided mentoring and instruction while preparing the course and video lectures. Joseph Freidman provided platform support for the first two offerings. Bob Iannucci contributed novel syllabus ideas. We have been extremely fortunate with our TAs, whose talents and contributions make this course possible. Finally, we thank the students who make teaching an exceedingly rewarding and worthwhile experience and who provide invaluable feedback that allows us to continuously improve the course.